\documentclass[aps,pra,reprint,showpacs,superscriptaddress,floatfix]{revtex4-2}
\usepackage{comment}
\usepackage{graphics,amssymb,amsmath,epsfig,color,textgreek}
\usepackage{graphicx}
\usepackage[dvipsnames]{xcolor}
\usepackage{dcolumn}% Align table columns on decimal point
\usepackage{bm}% bold math
\usepackage[colorlinks=true,citecolor=cyan]{hyperref}
\hypersetup{colorlinks=true,citecolor=cyan,linkcolor=red,urlcolor=magenta}
\usepackage{braket}
\usepackage[normalem]{ulem}
\usepackage{cancel}
\usepackage{diagbox}
\usepackage{lipsum}
\usepackage{algorithm}
\usepackage[noend]{algpseudocode}
\usepackage{cleveref}
\usepackage{orcidlink}
\usepackage[mathscr]{eucal}

\crefformat{figure}{Fig.~#2#1#3}
\crefformat{equation}{Eq.~#2#1#3}
\crefformat{appendix}{App.~#2#1#3}
\crefformat{section}{Sec.~#2#1#3}

\newcommand{\opX}{\mathbf{X}}
\newcommand{\opY}{\mathbf{Y}}
\newcommand{\opZ}{\mathbf{Z}}
\newcommand{\opA}{\mathbf{A}}
\newcommand{\opB}{\mathbf{B}}
\newcommand{\ham}{\mathbf{H}}

\newcommand{\Dchi}{D}

\newcommand{\NCSU}{Department of Physics and Astronomy, North Carolina State University, Raleigh, NC 27607, United States}

\newcommand{\IU}{Department of Chemistry, Indiana University, Bloomington, IN 47405, USA}

\begin{document}

\title{Efficient computation of real-time correlators using Pauli Propagation}

\author{Alexander~F.~Kemper\orcidlink{0000-0002-5426-5181}}
\affiliation{\NCSU}

\author{Raghav G. Jha\orcidlink{0000-0003-2933-0102}}
\affiliation{\NCSU}

\author{Arnab Bachhar\orcidlink{0009-0000-1170-0078}}
\affiliation{\IU}

\author{Goksu~C.~Toga\orcidlink{0000-0002-0316-2502}}
\affiliation{\NCSU}

\author{Mariano Guerrero Perez\orcidlink{0009-0002-1240-3432}}
\affiliation{\NCSU}

\author{Nicholas J. Mayhall\orcidlink{0000-0002-1312-9781}}
\affiliation{\IU}

\date{\today}

\begin{abstract}
Pauli propagation has shown promise for classically simulating quantum dynamics by evolving observables directly in the Heisenberg picture. In this work, we investigate its use for computing real-time two-point time-ordered correlators in one- and two-dimensional quantum systems. One major limitation of Pauli propagation is the rapid growth in the number of Pauli strings beyond short times. We overcome this limitation by combining accurate short-time Pauli-propagation data with time extension methods based on a positivity condition and the observation that the dynamics are often dominated by a small number of characteristic frequencies. This combined approach extends correlation functions far beyond the directly accessible time window while avoiding the exponential proliferation of Pauli operators. We demonstrate that the resulting correlators retain the relevant dynamical and spectral information. Our results broaden the regime in which classical methods can reliably probe the real-time dynamics of interacting quantum many-body systems.
\end{abstract}

\maketitle

\section{Introduction}

Quantum correlation functions, which encode the excitations out of the ground or thermal state, form the bridge between the microscopic details of the problem to the macroscopic observables.  In particular, the time-dependent correlation functions, which are of the form
\begin{align}
    C(t) = \mathrm{Tr} \left[ \rho \opA(t) \opB\right]
\end{align}
for some operators $\opA$ and $\opB$, are used to evaluate how a system responds to applied fields and as fundamental building blocks in many-body theory through the Green's function. The Fourier transform of correlation functions $C(\omega)$ are routinely measured in experiments such as neutron scattering, X-ray scattering, conductivity measurements, or optical spectroscopies. In short, they are the connective tissue between the theoretical foundation and the experimental observations.

As such, it is critical to be able to compute correlation functions efficiently; but despite decades of algorithmic improvements it remains a formidable task, in large part because performing the real time evolution of either the wavefunction or the operator is computationally expensive.  Performing the time evolution using traditional tools are typically limited: exact diagonalization rapidly hits the upper limits of available memory, real-time quantum Monte Carlo hits the phase problem, and tensor network methods do not scale efficiently beyond 1D due to sharp rise in the bond dimension during the evolution.

Yet from another perspective, the time evolution of the operator $\opA$, in particular for short times, should be relatively benign. There is a natural rate of spreading of the operator under the Heisenberg equation of motion and this is limited, i.e., its support does not immediately cover the entire system but rather grows over time.  In lattice systems this is captured by the Lieb-Robinson bound~\cite{Lieb1972, Hastings:2005pr}, which states that outside a light cone with a characteristic velocity, the operator support is exponentially suppressed.  Alternatively, from a computational perspective, first order discretized time evolution only involves a single commutator at each time step, limiting the spread of the operator over time.  This observation suggests that we could recast the evolution of the correlation function into finding the time-evolved operator $\opA(t) = \sum_i a_i P_i$ in some basis $\{P_i\}$, which hopefully remains modestly sized, and to evaluate the correlation function either directly or using the $P_i$ term by term.

The idea of computing $\opA(t)$ through the Heisenberg equation of motion has recently gained traction as a computational technique; it goes by the name Pauli propagation or Pauli path integrals, both of which evolve the original operator $\opA$ (often a Pauli string) growing through its operator space. 

\begin{figure*}[t]
    \centering
    \includegraphics[width=1.0\linewidth]{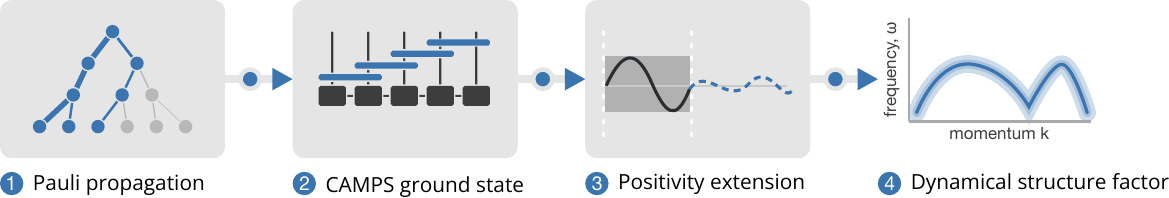}
    \caption{Overview of the computational strategy developed in this paper. 1) Real-time operator dynamics, truncated on coefficient and weight to control the growth of Pauli strings. 2) Clifford-augmented MPS supplies an accurate, low-bond-dimension approximation of the ground state to measure against. 3) A low-rank positive semi-definite (PSD) fit super-resolves short-time data out to late times, avoiding Pauli explosion. 4) By combining these techniques, we can obtain the dynamical spin structure factor $S(k,\omega)$}
    \label{fig:schema}
\end{figure*}

Pauli propagation grew out of efforts to classically simulate quantum circuits, where an operator is evolved in the Heisenberg picture and the resulting sum of Pauli strings is kept tractable by discarding high-weight contributions~\cite{Rall2019}. These methods drew sharp interest after the 2023 IBM Eagle experiment reported evidence for quantum utility in kicked Ising dynamics~\cite{kimEvidenceUtilityQuantum2023}, which motivated a series of classical calculations that matched or exceeded the quantum results~\cite{begusicFastClassicalSimulation2023, begusicFastConvergedClassical2024, tindallEfficientTensorNetwork2024a}. The same core idea, appearing under names such as Sparse Pauli Dynamics, Pauli backpropagation, and low weight efficient simulation algorithm (LOWESA)~\cite{Fontana:2023wvl}, now rests on a growing body of provable guarantees: polynomial-time simulation of noisy circuits~\cite{schusterPolynomialTimeClassicalAlgorithm2025} and noiseless average-case circuits~\cite{angrisaniClassicallyEstimatingObservables2025}, with further results for arbitrary local noise~\cite{angrisaniSimulatingQuantumCircuits2026}. Related to this current work, these techniques have recently been pushed from circuit benchmarking toward real-time operator evolution in two and three dimensions~\cite{Begusic:2024fty} and toward ground-state energy estimation~\cite{shrikhandePauliPropagationEnables2026}, showing that the reach of Pauli propagation extends well beyond its original setting. Additionally, Pauli shadow simulation approaches based on vectorization and inspired by classical shadows have been proposed as an efficient method of calculating time-ordered and out-of-time-ordered correlation functions on quantum computers. \cite{somma2025shadow,Chiew_2026srr,chakraborty2026efficient,Chiew_2026hsk}

While Pauli propagation is quite promising, it does not intrinsically avoid the exponential growth in the operator support over long times, requiring truncation which introduces biased errors. However, the growth (and thus error) can be mitigated when one computes local correlation functions for two reasons.  First, correlation functions in real systems are often bounded in time. Due to interactions both within the system and with some effective environment the system finds itself in, the correlation functions often decay. This process is reflected in the Fourier domain $C(\omega)$ as a finite linewidth $\tau^{-1}$ to the excitations that the correlation function probes.  And, on the time domain side, limits how far one has to propagate the operator $\opA$.  Second, correlation functions \textit{often} have a limited number of frequencies at which we see some signal, in particular for solids in momentum space (though this is not always true, see Sec. \ref{sec:1DAFM}). This suggests the use of low-rank decompositions in the time domain, which are further enabled by the fact that certain correlation functions are positive functions~\cite{kemper_denoising_2024}, limiting the number of time points needed even further~\cite{caratheodory1911variabilitatsbereich,toeplitz1911fourier,pisarenko1973retrieval}.

In this work, we combine Pauli propagation and short-time extensions (or approximations) to form an efficient algorithm for computing correlation functions, and we illustrate this on 1D and 2D Heisenberg models (see Fig.~\ref{fig:schema}). Due to the exponential growth of terms in Pauli propagation, one must choose a truncation scheme, and a few primary
options have been explored~\cite{begusicFastClassicalSimulation2023,rudolphPauliPropagationComputational2025}. Here, we show how truncation on the Pauli coefficient $\varepsilon$ and weight $w$ affects the resulting spectra and computational time. We also demonstrate how the low-rank nature of the signal can be exploited to produce high resolution spectra even from extremely limited time data ($\le 7$ time steps in the case of a 2D antiferromagnetic calculation).

One computational challenge unique to \textit{equilibrium} correlation functions is the requirement that we generally need to compute dynamical quantities with respect to true ground/thermal states (or approximations thereof). Beyond 1D, this complicates the direct use of DMRG ground states due to sharp rise in entanglement, which makes evaluating expectation values of large Pauli sums cumbersome. However, by recognizing that Clifford transformations can be used to significantly reduce an MPS bond dimension while simultaneously leaving the coefficient structure of a Pauli sum unchanged, we demonstrate that Clifford augmented matrix product state~\cite{Qian:2024vea} (CAMPS) can be used effectively for computing equilibrium correlation functions from Pauli Propagation trajectories.
We use this approach for computing the dynamical spin structure factor for the $8 \times 8$ anti-ferromagnetic Heisenberg lattice, which is at the edge of what can be handled with other methods.

Overall, we find that the combination of Pauli propagation and extension of data from short times yields a very promising tool for computing correlation function spectra (shown schematically in Fig.~\ref{fig:schema}). 

\section{Pauli propagation}

Pauli propagation (PP) is a classical method (originally developed for simulating quantum circuits~\cite{rallSimulationQubitQuantum2019}) for approximating the Heisenberg
evolution of an operator, and it is the engine we use to evaluate real-time
correlation functions. 

The central object is the time-dependent operator,
\begin{align}
    \opA(t) = e^{i\ham t}\, \opA\, e^{-i\ham t}
\end{align}
which we expand
in the basis of Pauli strings $P$ (Hermitian tensor products of $\{I,X,Y,Z\}$),
\begin{equation}
  \opA(t) = \sum_{i} a_i(t)\, P_i ,
  \label{eq:expansion}
\end{equation}
with real coefficients $a_i(t)$. This basis is natural in the Heisenberg picture:
any two Pauli strings either commute or anti-commute, their products are again
Pauli strings up to a phase, and the algebraic operations reduce to cheap binary
manipulations of bit masks.

Evolution proceeds by Trotterizing $e^{-i\ham t}$ into a sequence of small-angle Pauli rotations $e^{-i\theta P_k}$. Conjugating a string $P$ by such a rotation that leaves it
unchanged if $[P,P_k]=0$ and otherwise splits it into two terms,
\begin{equation}
  e^{i\theta P_k}\, P\, e^{-i\theta P_k}
   = \cos(2\theta)\, P + i\sin(2\theta)\, P_k P 
  \label{eq:branch}
\end{equation}
Each anti-commuting rotation therefore branches one string into two, generating a
binary tree whose leaves are the Pauli strings making up $\opA(t)$. Exact evolution
is exponentially expensive, but for small angles the $\sin$ branches are
suppressed, so the operator stays dominated by a comparatively small set of strings
with appreciable weight.

The cost is controlled by truncating this proliferation, which is done in two possible ways. First, after each step (single Pauli conjugation) we can discard
strings whose coefficient falls below a threshold, $|a_i| < \varepsilon$ (coefficient
cutoff). Second, we can drop strings whose Pauli weight, the number of non-identity
sites, exceeds a cap $w$ (weight cutoff). The coefficient cutoff removes the
dynamically negligible operators, while the weight cutoff directly limits how far
the operator can spread in operator space. Both are systematically convergent:
results become exact as $\varepsilon \to 0$ and $w \to N$, and the controlled trade-off
between accuracy and the retained number of Pauli strings to evaluate is what makes the method
efficient. 
Unlike weight truncation, which is a proper subspace projection, coefficient truncation has to be performed carefully because there is no well-defined continuous time limit for a given fixed $\varepsilon$, 
and the truncation threshold generally needs to decrease linearly with time step size
~\cite{Begusic:2024fty}. 
In practice, $\varepsilon$ and $w$ are chosen by convergence studies, and that limits the use of correlators to a time window in which the truncated dynamics remain
converged. 

\section{Computing correlation functions with Pauli propagation}

\begin{figure}[t]
    \centering
    \includegraphics[width=1.0\linewidth]{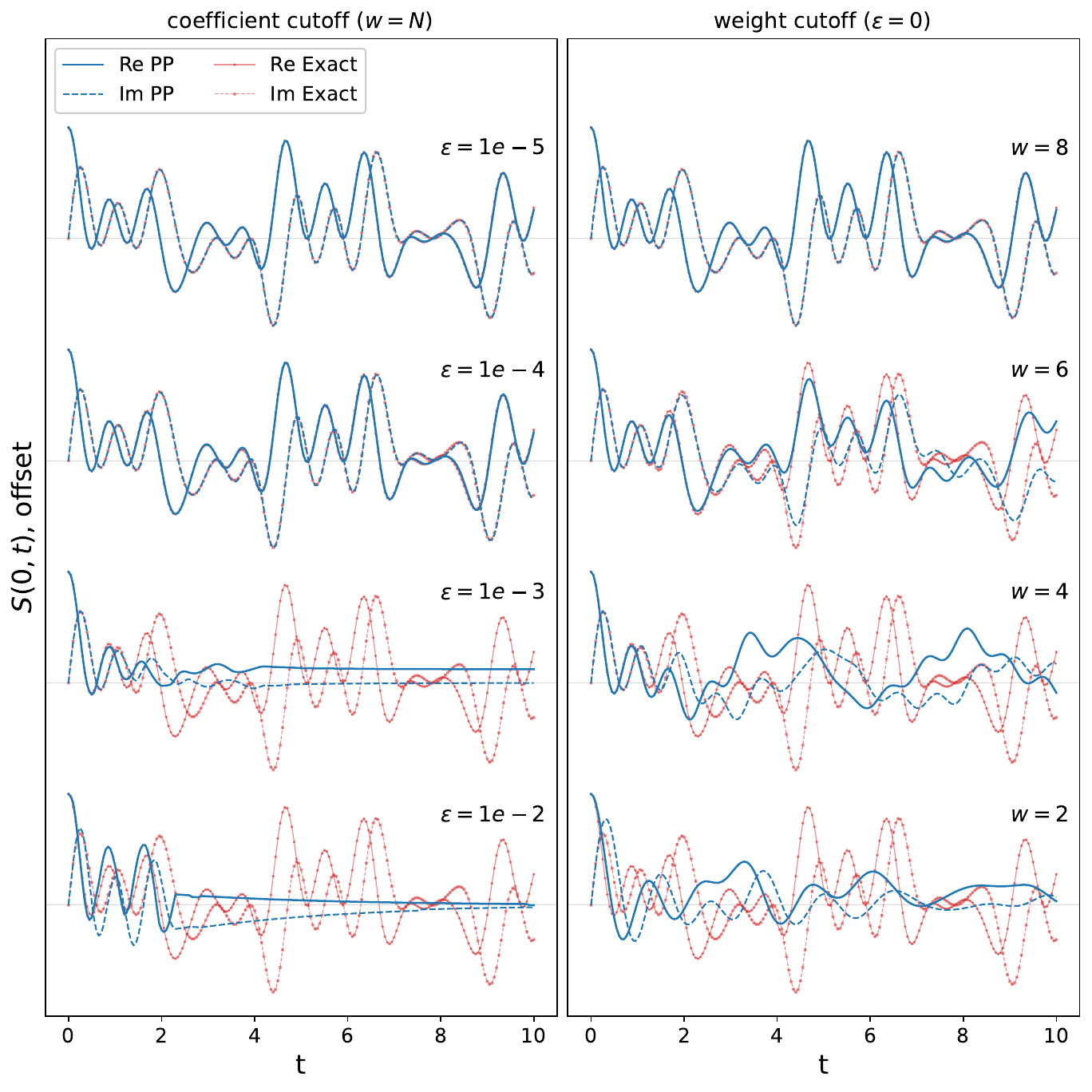}
    \caption{On-site correlation function $S(r=0,t)$ for the $N=8$ 1D
    ferromagnetic Heisenberg chain. Pauli propagation (PP) results are shown in lines, and the exact results with red markers.     }
    \label{fig:n8_autocorr}
\end{figure}

Correlation functions are of the form
\begin{align}
    C_{AB}(t,t') = \mathrm{Tr} \left[ \rho \opA(t) \opB(t') \right]
\end{align}
for some operators of interest $\opA$ and $\opB$.  Typically, we are interested in equilibrium correlation functions which only depend on the time difference $t-t'$, and we thus set $t'=0$ and work with $C_{AB}(t)$.  We have a problem that is naturally formulated in the language of Pauli propagation, i.e., the evaluation of $\opA(t) \opB$ with respect to some density matrix $\rho$.  Moreover, in materials and molecules, correlation functions are often \emph{damped}. This is most easily seen from the wavefunction picture. The correlation function for an eigenstate of the Hamiltonian $\ham$ involves the inner product between $\opA \ket{\psi}$ and $e^{-i\ham t}\opB \ket{\psi}$. The excitation $\opB \ket{\psi}$ spreads through state space, and as it does, some amplitude is lost to the effective environment that the system finds itself in.  In the language of many-body theory, this loss is captured in an effective self-energy for the operator, but more simply, it is some decaying function that appears as memory is lost to the environment.  For our purposes, what this does is limit the amount of time for which we need to simulate, increasing the effectiveness of Pauli propagation for such an approach.

In this section, we outline the basic uses of Pauli propagation for computing correlation functions. We will use the Heisenberg model as a test case
\begin{align}
    \ham = J \sum_{i=1}^{N-1} \opX_i \opX_{i+1} + \opY_i \opY_{i+1} + \opZ_i \opZ_{i+1},
    \label{eq:Heisenberg_Hamiltonian}
\end{align}
where $\opX_i/\opY_i/\opZ_i$ are the Pauli operators acting on site $i$. We set $J = \pm 1$ for anti-ferromagnetic and ferromagnetic case respectively, and use $J$ as our energy unit. We note that although the ground state complexity strongly depends on the sign of $J$, the ferromagnetic and anti-ferromagnetic regimes are equivalently complex from the operator perspective. Throughout, we will be calculating a transverse spin-spin correlation function ($\opA = \opB = \opX_{i/j}$ for site indices $i/j$), and thus we will specialize the notation accordingly and use $S(r,t)$ for the dynamical spin structure factor.

As mentioned before, we will investigate two independent cutoff strategies (on coefficient $< \varepsilon$, and on Pauli weight $\le w$) on a small system where the calculation can be performed. 
Here, and for the remainder of the discussion, we use a Trotter time step $\Delta t = 0.05$.
Fig.~\ref{fig:n8_autocorr} shows the on-site 
spin-spin correlation function $S(r=0,t) = \langle \Psi_{\text{FM}} \vert \opX_0(t) \opX_0 \vert \Psi_{\text{FM}} \rangle$.
The coefficient cutoff and weight cutoff affect $S(r=0,t)$, but in different ways.  The coefficient truncation yields essentially exact results for $\varepsilon=10^{-4}$ and below; above, the initial oscillations are correct but it rapidly diverges from the exact solution and loses all dynamical information.  The weight cutoff also shows initial agreement, but devolves even with a relatively modest weight cutoff ($w=6$).  The peak Pauli counts for the two cases where there is reasonable agreement are $16,368$ for $\varepsilon<10^{-4}$ and $10,368$ for $w \le 6$; relatively high fractions of the full operator space which has $65,535$ elements or the symmetry-allowed operator space which has $16,384$ elements. Although this is not a particularly large operator space as here $N=8$, this situation gets worse for larger $N$.

\begin{figure}[htpb]
    \centering
    \includegraphics[width=1.0\linewidth,clip=true, trim=0 0 0 0]{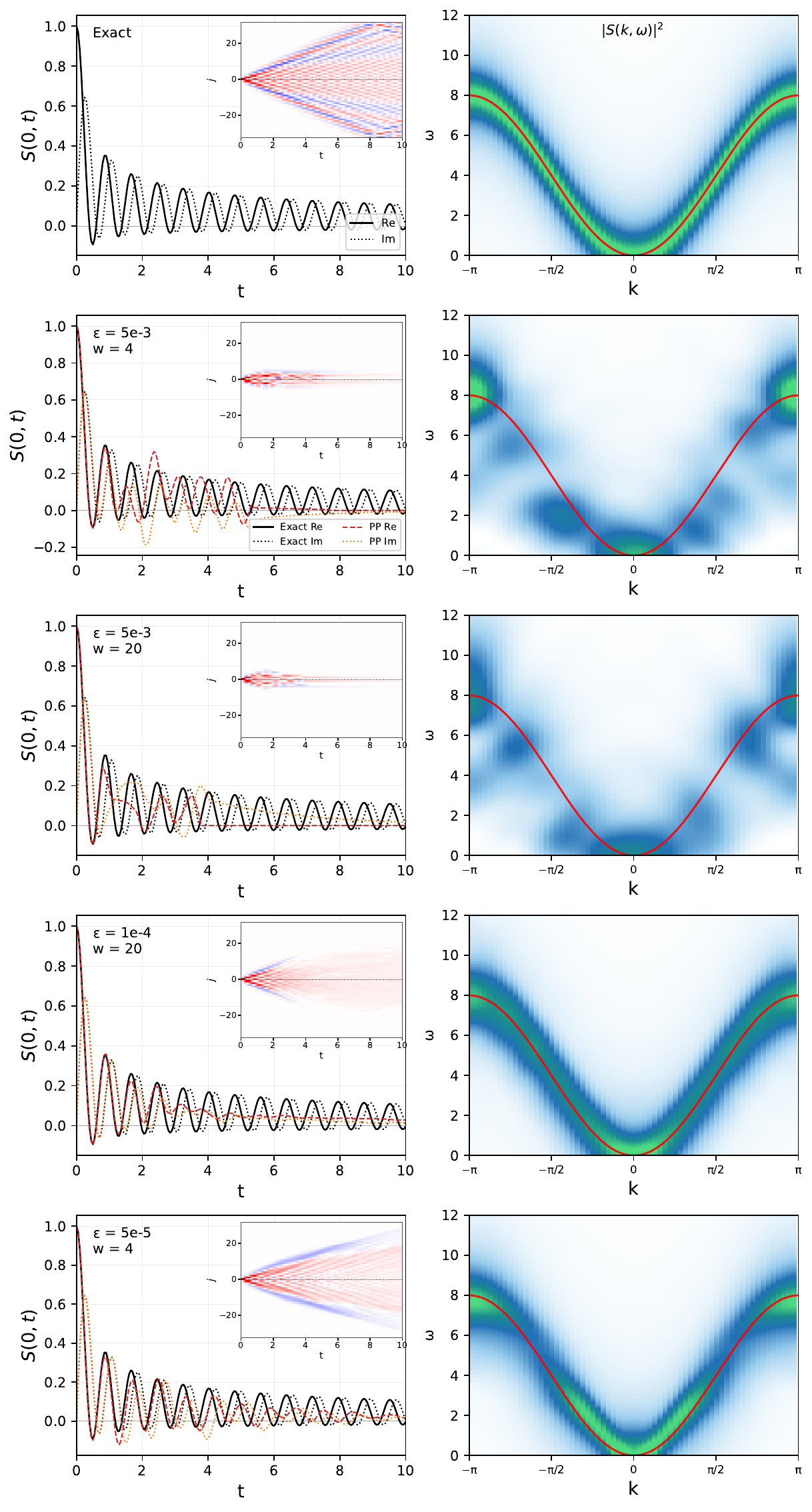}
    \caption{Pauli propagation convergence overview for the 1D ferromagnetic Heisenberg
    chain at $N=64$. The top row is
    the exact (analytic) 1-magnon reference; the remaining rows are four
    representative composite-truncation points $(\varepsilon, w)$.
    \emph{Left column}: On-site correlation function $S(r=0,t)$, real (solid) and
    imaginary (dotted) parts, with ED in black and PP overlaid in
    red/orange. Insets show $\operatorname{Re} S(r=j,t)$ as a
    function of time (horizontal) and site $j$ from the
    source (at $r=0$). \emph{Right column}: dynamical structure factor
    $|S(k,\omega)|^2$ with exponential time-domain
    damping $e^{-t/\tau}$ using $\tau=1$. The cyan curve marks the analytic
    1-magnon dispersion $\omega(k) = 4(1-\cos k)$. Each momentum is individually normalized.
    }
    \label{fig:n64_fm_overview}
\end{figure}

\subsection{Damped correlation functions}

The early-time agreement, even for more relaxed cutoffs, suggests an alternative approach: limit the consideration to early times and work with that data. This is supported by the discussion above on the damping of correlation functions. To obtain the spectra, we can apply an ad hoc damping function $\exp(-t/\tau)$ that ensures the major contribution to the spectrum comes from the time regions where the Pauli propagation agrees with the exact results.  This approach has been used before in a variety of contexts~\cite{lee2022compact,kokcu2024linear}, typically with the purpose of limiting the amount of data needed.  Its effect on the spectrum is somewhat modest, it results in the convolution of the spectrum with a Lorentzian of width $\tau^{-1}$. In order to be able to correctly obtain frequencies within that width, $S(r,t)$ should be accurate to roughly $t=3\tau$, where the exponential damping is effectively zero.

\subsubsection{Heisenberg Ferromagnet in 1D}

We apply this idea to the 1D Heisenberg model, now with $N=64$ sites, and consider coefficient cutoff, weight cutoff, as well as their combination. 
At this system size, it rapidly becomes prohibitively expensive to work without any cutoff in terms of both the maximal number of Paulis and cumulative time taken for the calculation out to $t_{\text{max}}=10$ (we present maximum Pauli count and wall time considerations in App.~\ref{app:walltime}).

In Fig.~\ref{fig:n64_fm_overview} we show the resulting correlation functions for a few selected combinations of coefficient and weight truncation. The time traces of the two best cases show reasonable agreement with the exact reference, but only qualitatively. The two worst cases show little resemblance to the reference.  Interestingly, however, when applying the ad hoc damping function and Fourier transforming in space and time,
all four cases produce a spectrum that contains the features of the known spectrum: a smoothly dispersing magnon band.  

In the case of the more relaxed cutoffs, the spectrum is rather distorted, but as the accuracy of the Pauli propagation is increased, the agreement becomes much better.  Importantly, the agreement at the level of the spectrum did not require absolute agreement with the reference; the time traces for the best cases only agree qualitatively, even early on.

One important point to note is that the individual momenta are not equally prominent in the spectrum. The frequencies that appear are correct, and to illustrate this point the momenta are individually normalized.

\subsubsection{Heisenberg Anti-Ferromagnet in 1D}
\label{sec:1DAFM}

The observations above carry over in the anti-ferromagnetic regime as well. The notable distinction here is that, unlike the previous case of the ferromagnetic Heisenberg model, where the ground state is a product state, the ground state of the anti-ferromagnetic Heisenberg model is more complex. The first step is to prepare the ground state for both trajectory techniques we study (TDVP and PP). 
To do this, we use standard DMRG to efficiently find an accurate ground state, which for N=64 is within 1\% of the exact ground state energy in the thermodynamic limit~\cite{Hulthen1938}
i.e., $E_0/N=1-4 \ln(2)$ with the definition of the Hamiltonian used in Eq.~\eqref{eq:Heisenberg_Hamiltonian}.  For this MPS ground state calculation we set the maximum bond dimension $\chi_{\rm max}=400$.

\begin{figure}[b]
    \includegraphics[width=0.49\textwidth,clip=true, trim=0 0 0 20]{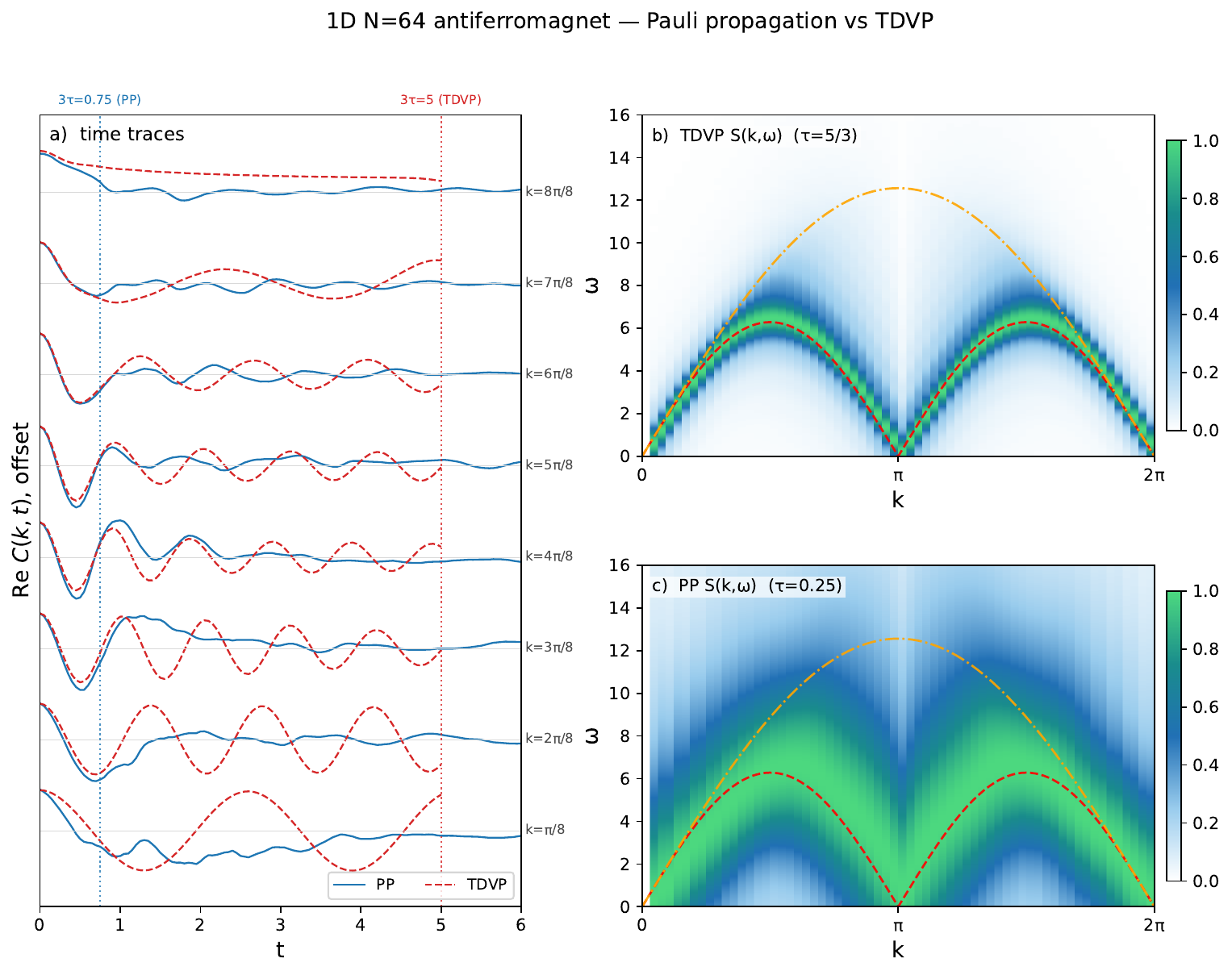}
    \caption{Pauli propagation results for the 1D Heisenberg anti-ferromagnet. a) Time traces for selected momenta using Pauli propagation (PP) and TDVP. b,c) $|S(k,\omega)|^2$ obtained (b) from TDVP with $\tau=5/3$ and (c) from Pauli propagation with $\tau=0.25$. The red and orange dashed lines indicate the known boundaries of the spectrum of spinons in 1D.}
    \label{fig:1dafm}
\end{figure}

Because the ground state is no longer a product state for the AF Hamiltonian, the evaluation of Pauli expectation values is no longer computationally negligible, and actually ends up becoming a computational bottleneck.  After solving for our ground state with DMRG, we can compute the dynamics with either Pauli propagation or with standard TNS methods, such as TEBD or TDVP. In Fig.~\ref{fig:1dafm} we present a comparison between Pauli propagation and direct Schr\"odinger picture evolution of the MPS ground state via TDVP.  
We have performed convergence studies to determine that a bond dimension of $\chi=400$ is suitable for obtaining converged MPS-TDVP results for this 1D system, providing a reliable ground truth for benchmarking the performance of the PP-derived correlation functions. 
To limit the computational cost, we use a very loose PP truncation ($\varepsilon = 10^{-3}$ and simulate up to $t=10$ with $\Delta t=0.05$.
Because of this aggressive truncation, the size of the Heisenberg operator never exceeds 5968 Paulis.
While evolving the operator with tighter thresholds is certainly tractable, the computational limitation here is the evaluation of the expectation value, $\bra{0}X(t)X(0)\ket{0}$ when $\ket{0}$ is an MPS wavefunction. Since the computational bottleneck is just the expectation value sampling, one could alternatively use a tighter truncation for the dynamics, and then approximate the expectation values (see Appendix \ref{app:improved_afm}), however we leave a detailed comparison of this approach to future work.

In Fig. \ref{fig:1dafm}, the aggressive PP truncation effects are clearly visible in both the time traces as well as the spectrum. 
In the time domain, the PP results appear accurate only to very short times.
This necessitates a relatively short damping cutoff $\tau=0.25$, creating significant frequency broadening. 
Despite this, the Pauli propagation spectrum still shows the correct generic features of the anti-ferromagnet, a strongly dispersive mode which breaks into a continuum of spinon excitations~\cite{Kulka2025} near the peak of the band, but the resolution is not at the same level of accuracy as the MPS-TDVP results. While one would not expect PP to out-perform an MPS simulation in 1D, 
the current results nicely highlight the challenges that arise for entangled ground states. We leave a more detailed study of how the coefficient, weight, and mixed truncation of PP affect the spectrum at high symmetry points of the Brillouin zone for future work.

\subsection{Extracting spectra with positive semi-definite extension}

The results above point to a potential path towards the use of Pauli propagation for computing correlation functions: limit the consideration to early times, when the agreement is still reasonable, and make as much of that as possible. In the previous sections, we accomplished this by an ad hoc (but physically inspired) damping function. While the use of this exponential windowing function helps concentrate the FFT onto the earlier (more accurate) signal, standard Fourier theory provide limits to how short the signals can be while not losing all information. However, when the spectrum consists of a limited number of frequencies (as in the case of the ferromagnet), further advances based on low-rank decompositions are possible.

Recently, it was shown\cite{kemper_denoising_2024} by one of us that correlation functions of the form
\begin{align}
    C_{AA}(t,t') = \mathrm{Tr} \left[ \rho \opA^\dagger(t) \opA(t') \right]
\end{align}
are \emph{positive} functions. Equivalently, when their time-discretized versions are arranged as a (Toeplitz) matrix, this
matrix is positive semi-definite (PSD).  In Ref.~\cite{kemper_denoising_2024} this was used to remove noise from correlation functions and, as is relevant to the work here, extended to longer times from initial data. 
This relies on the fact that a rank-$r$ PSD Toeplitz matrix can be uniquely determined from $2r$ noiseless numbers per the foundational work of Carath\'eodory\cite{caratheodory1911variabilitatsbereich},  Toeplitz\cite{toeplitz1911fourier}, and Pisarenko\cite{pisarenko1973retrieval}. In other words the matrix admits a unique (Vandermonde) decomposition in terms of a few ($r$) oscillatory components\cite{yang2018frequency}, and knowledge of these oscillatory components can then be trivially used to extend the matrix. In the case of noisy data, a similar approach can be used to extend the matrix in such a way that it maintains both positivity and its Toeplitz structure~\cite{kemper_denoising_2024}.

\begin{figure}[t]
    \centering
    \includegraphics[clip=true,trim=0 0 0 0,width=1.0\linewidth]{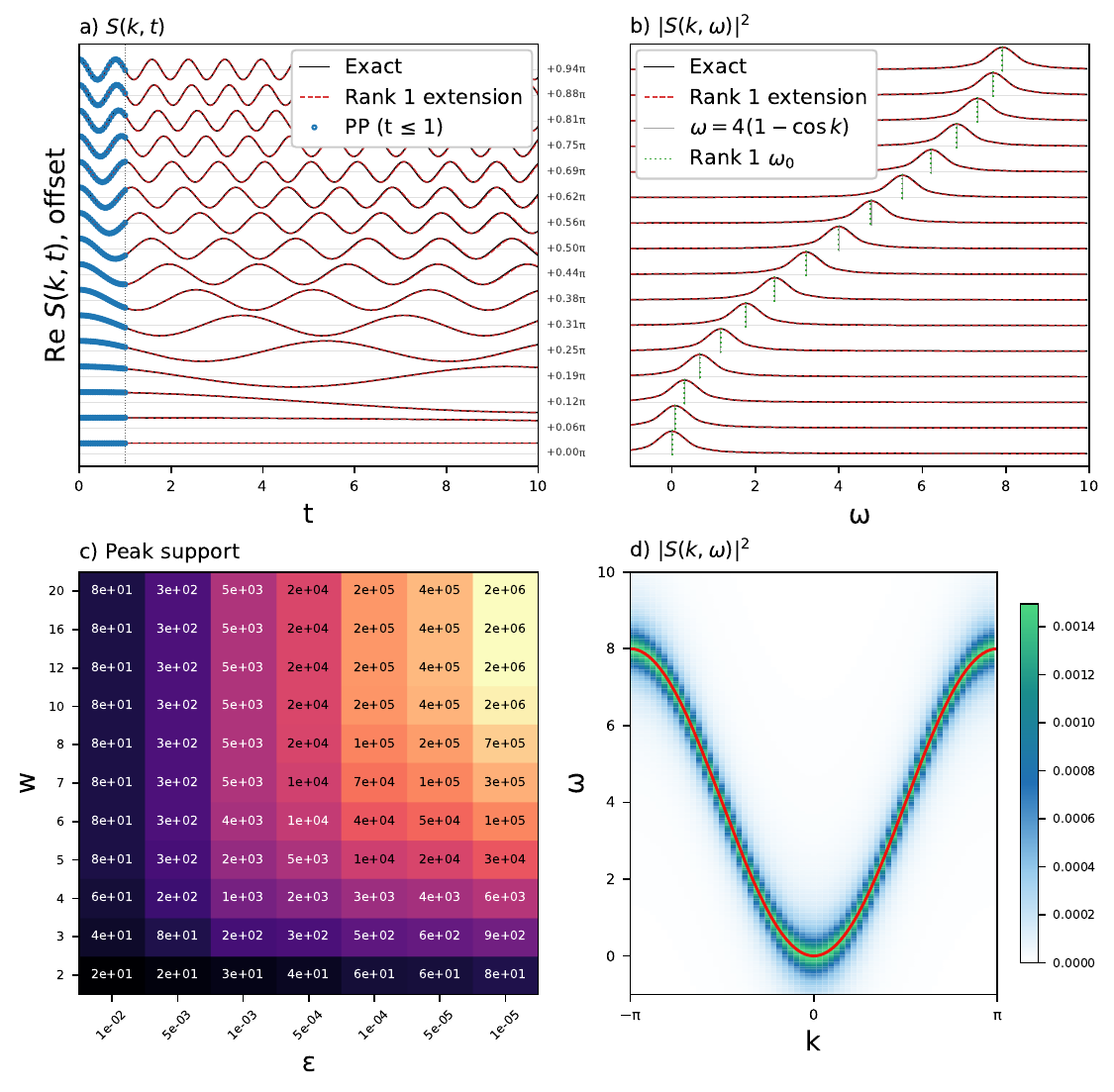}
    \caption{
    Positive semi-definite extension in momentum space from $N=64$ ferromagnetic Heisenberg in 1D for short-time
    Pauli Propagation (PP) data ($\varepsilon = 10^{-5}$, $w = 20$, 21 samples).
    a) $\operatorname{Re} S(k,t)$ showing 16 momenta, comparing exact (ED) with the rank 1 extension from PP data (blue markers). Vertical lines indicate the rank 1 frequency found ($\omega_0$) as well as the exact analytic result at each momentum $k$.
    b) Resulting power spectrum of $S(k,\omega)$ with exponential damping using $\tau=2.5$ and zero-padding.
    c) Maximal operator support (number of Pauli terms) during the propagation.
    d) Full $S(k,\omega)$ heatmap
    from the PSD-extended signal.}
    \label{fig:n64_fm_psd_qspace_short}
\end{figure}

Here, we use this method to correct some of the errors associated with the truncations in Pauli propagation and as a super-resolution method to extract the spectrum from short time data. The correlation function in the spatial Fourier domain has $\opA=\opX_k$, and the theorem from Ref.~\cite{kemper_denoising_2024} applies. Moreover, in momentum space, the number of frequencies (equiv. matrix rank $r$) is often small, which makes this type of method particularly attractive.
We demonstrate how this works in Fig.~\ref{fig:n64_fm_psd_qspace_short}a). The Pauli propagation calculation is run out to $t=1$ ($21$ data points), and the remaining data is produced from the PSD extension.  For most momenta the extension is nearly identical to the exact results; some deviations appear around $k=\pi/2$ which are due to the fact that Pauli propagation is not exact.
It is not surprising then that the spectrum (here also calculated with a damping using $\tau=2.5$ purely for plotting reasons) agrees with the exact result.  The figure also shows that the peak number of Paulis for small cutoffs ($\varepsilon,w$) is somewhat smaller compared to the much longer runs shown in Fig.~\ref{fig:n64_fm_cost_heatmap}. For the smaller cutoff, limiting $t_{\text{max}}=1$ is what enabled the short time runs to complete at all.

In summary, making use of a physically-informed super-resolution method further improves the capabilities of Pauli propagation, overcoming some of the limitations imposed by the extremely rapid growth in operator space.

\subsection{2D Heisenberg Anti-Ferromagnet}
\label{sec:2d}

We next demonstrate the use of Pauli propagation for computing correlation functions in a two-dimensional anti-ferromagnetic Heisenberg system, which is a much more complicated problem. First, because of the higher connectivity in a two-dimensional lattice, the number of Pauli strings that are generated at each time step grows much more rapidly. Second, the 2D geometry requires much larger bond dimension for the ground state MPS. 
Obtaining a ground state for any non-trivial model is difficult, as traditional methods such as exact diagonalization and MPS-based techniques run into computational limitations.
The first issue can be overcome by the PSD extension, which both corrects some of the errors that occur due to the truncated Pauli propagation, and severely reduces the number of time points that need to be calculated.  The second issue is somewhat inherent to the system under consideration, and cannot be easily avoided; however it can be somewhat ameliorated. To overcome the steep rise of bond dimension required for preparing ground states of two-dimensional systems, we make use of a recently developed method that augments the MPS approach with Clifford circuits. This Clifford-augmented MPS (CAMPS)~\cite{Qian:2024vea} approach leads to substantial reduction in the required bond dimension for a fixed accuracy of the ground state (see Appendix~\ref{app:camps}), enabling the study of an $8\times 8$ Heisenberg model in the anti-ferromagnetic state. Because the computation of expectation values of Pauli sums can easily become a computational bottleneck, the CAMPS approach is uniquely well-suited for our task because the associated Clifford circuits can just be applied directly to the Heisenberg operators without increasing the number of Pauli strings.

\begin{figure}[t]
    \includegraphics[width=0.48\textwidth]{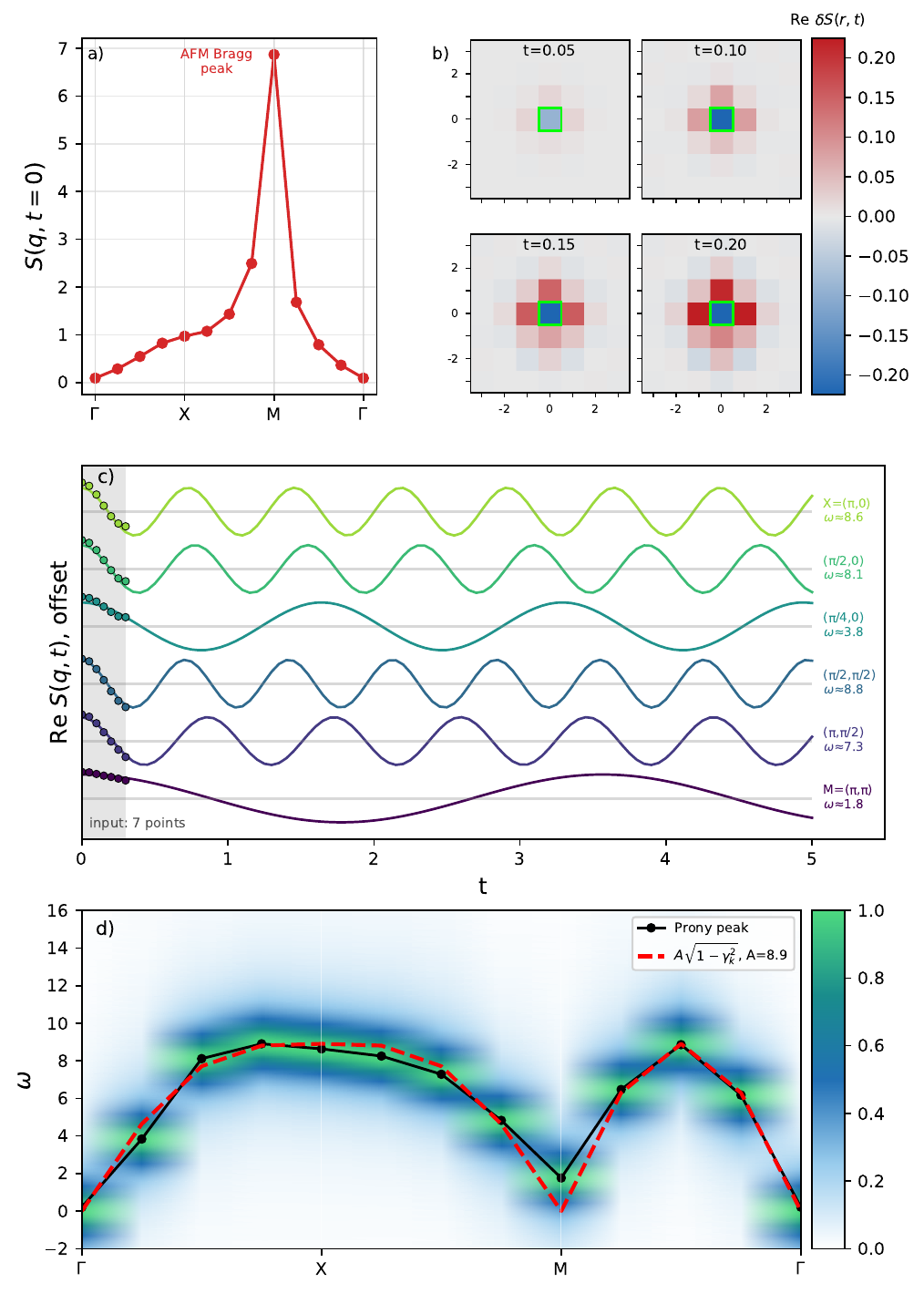}
    \caption{2D anti-ferromagnetic Heisenberg spectrum from Pauli propagation with $(w,\varepsilon)=(5,10^{-4})$ and CAMPS with $D=240$. 
    a) Static order parameter showing the presence of anti-ferromagnetic order.
    b) Spreading of the excitation $\delta S(r,t) = S(r,t) - S(r,0)$ for the time points used.
    c) Early time data from Pauli propagation and the rank-1 PSD extension for selected momenta.
    d) Spin structure factor $|S(k,\omega)|^2$ from the extended spectra, extracted frequencies from the rank-1 decomposition, and a fit to the linear spin wave spectrum $\omega_k = A\sqrt{1-\gamma_k^2}$.
    }
    \label{fig:2D_AF}
\end{figure}

Fig.~\ref{fig:2D_AF}(a) shows the static structure factor $S(k,t=0)$ along the high symmetry lines in the Brillouin zone. The clear peak at $M=(\pi,\pi)$, indicating a checkerboard pattern in the static spin structure, shows it is in the anti-ferromagnetic phase, as expected~\cite{Schulz1992}. 
We next evaluate the dynamical structure factor $S(r,t)$ using Pauli propagation, and show the change from the unperturbed state in Fig.~\ref{fig:2D_AF}(b). $S(r,t)$ decreases at the perturbation site (outlined in green), and subsequently spreads out through the lattice. The spread is limited because the time evolution is short, which would na\"ively indicate a severe limitation on resolving the dynamics. We overcome this limitation with rank-1 PSD extension, shown in Fig.~\ref{fig:2D_AF}(c). With the 7 time points calculated with Pauli propagation and CAMPS (in the gray box), we compute a rank-1 extension out to $t=5$ (recall that this time is arbitrary, as the rank-$r$ Vandermonde decomposition already tells us the frequencies). 
The full spectrum of $S(k,\omega)$, shown in panel (d), agrees well with the predictions of linear spin wave theory which says the peaks in the spectrum should appear at $\omega=A\sqrt{1-\gamma_k^2}$, where $\gamma_k = (\cos k_x + \cos k_y)/2$. There is a Goldstone mode at $\Gamma$, a rapid rise of the dispersion out to $X$, and then a softening at $M$.  Due to finite size effects, the soft mode at $M$ is not exactly at 0. In Appendix~\ref{app:improved_afm}, we discuss finite size scaling, and further demonstrate the agreement with linear spin-wave theory.

One observation about these results is that the limited spreading of the operators indicates that the Pauli propagation results can in principle be used for a larger lattice as well.  Similar to the 1D Heisenberg model discussed above, the limitation in size comes from the need to prepare the ground state and evaluate the Paulis, rather than the unbounded operator growth that occurs at later times than we are considering here. We demonstrate this in App.~\ref{app:largeFM} for a $100 \times 100$ ferromagnetic lattice, where we fully recover the known magnon spectrum at that lattice size.

\section{Discussion}

In this work, we show that Pauli propagation can provide a way to compute correlation functions from the perspective of the Heisenberg evolution of the operators.  The early time evolution captures the overall response reasonably well, and for single-site excitations avoids the need for time evolution under a Hamiltonian matrix in the full Hilbert space.  This enables finding correlation functions for very large system sizes at minimal cost, in particular when the state of interest is easily prepared, e.g., a computational basis state or some weakly entangled state. 
The approach is quite general, any method that can produce expectation values of Pauli strings in an efficient manner could make use of the Pauli propagated operators to calculate correlation functions. This could include tensor network methods, quantum Monte Carlo, and using a quantum computer directly or indirectly via sample Quantum Krylov diagonalization~\cite{Yu:2025czp}.

There are, however, some clear constraints, most notably that the number of Pauli strings grows extremely rapidly during the Heisenberg evolution. Formally, the number of such terms covers the entire dynamical Lie algebra after any finite time; the dynamical Lie algebra is exponentially scaling for most (if not all) problems of 
interest~\cite{kokcu2024classification,wiersema2024classification}, and it is only the finite time steps and the truncation that avoids the immediate explosion. The calculations above show that aggressive truncation does reasonably well in capturing the dynamics, but nevertheless the number of terms grows quite rapidly with time. 

The low rank PSD factorization, which is akin to a few-frequency Prony fit, can help with extracting the maximum amount of information, correcting some of the errors that occur due to the truncated Pauli propagation, and (uniquely) extend the results to longer time.  As shown above this can work quite well, but there are underlying assumptions that (a) the number of distinct frequencies is limited and (b) the data captures the dynamics sufficiently accurately that the frequencies can be clearly identified. The first assumption fails, for example, when dealing with a continuum of frequencies such as the spinon excitations found in the 1D anti-ferromagnet (see Sec.~\ref{sec:1DAFM}).  The second requires the Pauli propagation to be sufficiently accurately calculated, i.e., for the truncation to be not too severe.  The minimally acceptable truncation is an open question, which we defer to a future work.

It should also be noted that the current state of the art for calculating dynamics in 1D based on tensor networks is sufficiently good that it is not expected that Pauli propagation could ever be competitive, in particular where the low-rank assumption fails.  However, the situation is more interesting in higher dimensions where the efficiency of tensor state network methods is greatly reduced.  We illustrated this in Sec.~\ref{sec:2d} for the 2D anti-ferromagnet, where by combining PP dynamics with a ground state calculated using CAMPS~\cite{Qian:2024vea} we were able to obtain reasonably accurate spectra for a problem where TNS calculations become too expensive. 

These results suggest that the combination of (i) Pauli propagation dynamics, (ii) appropriate ground state approximations, and (iii) efficient signal processing techniques provides a promising path forward to go beyond the current state of the art for computing dynamical correlation functions on large, high-dimensional lattices.

\begin{acknowledgments}
We acknowledge support by the U.S. Department of Energy, Office of Science, Office of Advanced Scientific Computing Research under Award Number DE-SC0025430.
The CAMPS calculations (R.G.J) were supported by the U.S. Department of
Energy, Advanced Scientific Computing Research, under
contract number DE-SC0025384.
The Pauli propagation used the PauliOperators.jl package which was supported by the National Science Foundation (Award No. 2414574). 
\end{acknowledgments}

\bibliography{kemperlab,ref,nick}

\appendix

\renewcommand\thefigure{S\arabic{figure}}  
\renewcommand\thetable{S\arabic{table}}  
\setcounter{figure}{0}

\appendix

\section{Pauli Propagation with Clifford augmented matrix product states (CAMPS)}
\label{app:camps}

We need a robust and cost-effective method of computing the ground states of both 1D and 2D AFM Heisenberg model to compute the dynamical correlation function: 
\begin{equation}\label{eq:Cjt}
  C(j,t) = \bra{\Psi_0} \opX_c(t)\, \opX_j(0) \ket{\Psi_0}\,,
\end{equation}
where $\ket{\Psi_0}$ is the ground state of the lattice Hamiltonian, $c$ is a fixed
center site, and $j$ runs over all lattice sites.  The Heisenberg
time evolution is: 
$\opX_c(t)=e^{i\ham t}\,\opX_c\,e^{-i\ham t}$.  Our approach combines Pauli Propagation~(PP), which represents $\opX_c(t)$ as a sum of
Pauli strings with numerically computed coefficients, and
Clifford-Augmented MPS~(CAMPS)~\cite{Qian:2024vea} which represents $\ket{\Psi_0}$
as a Clifford-dressed matrix product state.
CAMPS is a refinement of MPS approach to ground state preparation and was shown to 
reduce the Clifford-based entanglement and improve the accuracy 
of standard MPS method. Due to its effectiveness, CAMPS has been combined with TDVP~\cite{Haegeman:2011zz} 
for time evolution~\cite{Mello:2024jev, Qian:2024wun},
finite-temperature simulations~\cite{Qian:2024lqx}, and 
for studying fermionic systems~\cite{Huang:2024ron} classically. 
In 2d, where the bond dimensions can grow substantially, this has been shown to be very effective~\cite{Li:2025mbw}. We use 
CAMPS to lower the Clifford-based entanglement and attain fixed accuracy with smaller bond dimension. This leads to significant reduction in computation of correlators using PP. The CAMPS ansatz proposes to write the ground state as: 
\begin{equation}
  \ket{\Psi_0} \;=\; \mathcal{C}^\dagger \ket{\psi}\,,
\end{equation}
where $\mathcal{C}$ is a \emph{Clifford unitary}; a product of two-qubit Clifford
gates found by an iterative optimization and $\ket{\psi}$ is a
conventional (standard) MPS of bond dimension~$D$. Since Clifford gates map Pauli operators to Pauli operators (up to phases), so conjugating any Pauli
string $P$ gives another single Pauli string:
$\mathcal{C}\,P\,\mathcal{C}^\dagger = e^{i\phi}\,P'$ with some phase factor. Therefore, the Hamiltonian which is built out of Paulis can be transformed to Clifford frame without much overhead. In particular, CAMPS optimization alternates:
(i)~select the best Clifford gate from a library and append it to $\mathcal{C}$;
(ii)~re-optimize $\ket{\psi}$ via DMRG in the new Clifford frame. 
Each iteration lowers the entanglement that $\ket{\psi}$ must carry, allowing a smaller $\Dchi$ to achieve a given energy accuracy.

\section{Improved truncation for 2D anti-ferromagnet}
\label{app:improved_afm}

In Sec.~\ref{sec:2d} we presented the $8\times8$ anti-ferromagnetic Heisenberg result obtained from a single fixed combination of ground-state bond dimension ($D$) and Pauli-sum truncation.
Here we show that this result can be improved by using a better MPS ground state, obtained from a larger-bond-dimension DMRG calculation.
However, evaluating the correlation function directly with such a large-bond-dimension ground state (contracted with large evolved sum of Paulis) is computationally expensive, so we introduce a controlled approximation and examine how well it performs.

We evaluate the correlation function of Sec.~\ref{sec:2d} for the periodic $8\times8$ square-lattice Heisenberg antiferromagnet, $H=\sum_{\langle ij\rangle}(X_iX_j+Y_iY_j+Z_iZ_j)$, mapped onto a one-dimensional snake ordering for tensor-network manipulation.
Starting from a random MPS of bond dimension $D=100$, we converge the DMRG
ground-state calculation by systematically increasing the bond dimension to a final $D_{\rm opt}=2000$ (SVD cutoff $10^{-10}$, with a small subspace-expansion noise term to avoid local minima), yielding the reference wavefunction $|\Psi_f\rangle$.
A single-site Pauli operator $X_1(t)$ is evolved in the Heisenberg
picture via Trotterized Pauli propagation\cite{mayhallPauliOperatorsJl2023}, with the Pauli sum truncated after every elementary rotation using the composite coefficient-and-weight rule of Sec.~II: strings are discarded once their coefficient magnitude falls below $\varepsilon$ or their Pauli weight exceeds $w$.
Two truncation combinations are used to bound the propagation error, $(w,\varepsilon)=(6,10^{-6})$ and $(4,10^{-8})$ along with a Trotter time step $\Delta t = 0.05$ so that we can identify a comparable $(w,\varepsilon)$ truncation combination in which the number of surviving terms stays under control, reflecting the complementary weight/coefficient trade-off.

The simple straightforward approach is to convert the evolved operator $X_1(t)$ into an MPO and contract it directly with the optimized ground-state MPS,
$|\Psi_f\rangle$.
However, because the Pauli sum can contain a very large number of terms (sometimes in millions), the resulting MPO bond dimension can itself become very
large, making contraction with the $D=2000$ MPS both time-consuming and
memory-intensive.
We therefore adopt a working assumption that, provided a bond-dimension-truncated copy of $|\Psi_f\rangle$ retains a sizable overlap (close to 99\%) with $|\Psi_f\rangle$, it can be used in its place to approximate the correlation function.
A similar issue arises from MPO bond dimension itself: rather than truncating MPO directly and tracking its discarded weight, we instead create a copy of the operator at a given time step and truncate it directly before MPO construction, which again yields an approximate correlation function. However, this additional Pauli sum truncation does not affect the later time dynamics, as it is  only used to evaluate the correlator for that time step. 

Because $X_1(t)$ and $|\Psi_f\rangle$ each carry their own, independently controllable truncation error, we separate the two in a second, read-out-stage of truncation applied only at analysis time.
First, the reference MPS $|\Psi_f\rangle$ ($D_{\rm opt}=2000$) is SVD-truncated (no re-optimization) to smaller bond dimensions $D\in\{500,800,1200\}$, giving truncated states $|\psi_D\rangle$.
Second, for each $D$, $X_1(t)$ is further clipped at coefficient thresholds $\varepsilon_{\rm clip}\in\{5\times10^{-3},2\times10^{-3},1\times10^{-3}\}$; the discarded weight $\delta(\varepsilon_{\rm clip})=\lVert X_1(t)\rVert_2-\lVert X_{1,{\rm clip}}(t)\rVert_2$ (the $\ell_2$ norm of the dropped Pauli coefficients) is recorded alongside the surviving term count.

We then transform the truncated Pauli sum into an MPO via ITensor's~\cite{fishmanITensorSoftwareLibrary2020}
\texttt{OpSum} construction, and evaluate the expectation value directly against the truncated MPS, $\langle\psi_D|\,\mathrm{MPO}[X_{1,{\rm clip}}(t)]\,|\psi_D\rangle$.
Because $\delta$ is available for every $(D,\varepsilon_{\rm clip})$ pair, we fit the resulting expectation values against $\delta$ at fixed $D$ and extrapolate to $\delta\to0$, removing the coefficient-truncation bias and isolating the residual error due to the finite MPS bond dimension alone.
Repeating this extrapolation independently for $D=500,\,800,\,1200$ then
exposes the bond-dimension convergence.

Preliminary results indicate that aggressive truncation of both the MPS bond dimension and the Pauli sum can be tolerated.
However, determining precisely how loose a truncation still yields an accurate spectrum requires further investigation, which we leave to future work.

\begin{figure}[htpb]
    %\centering
    \includegraphics[width=0.48\textwidth]{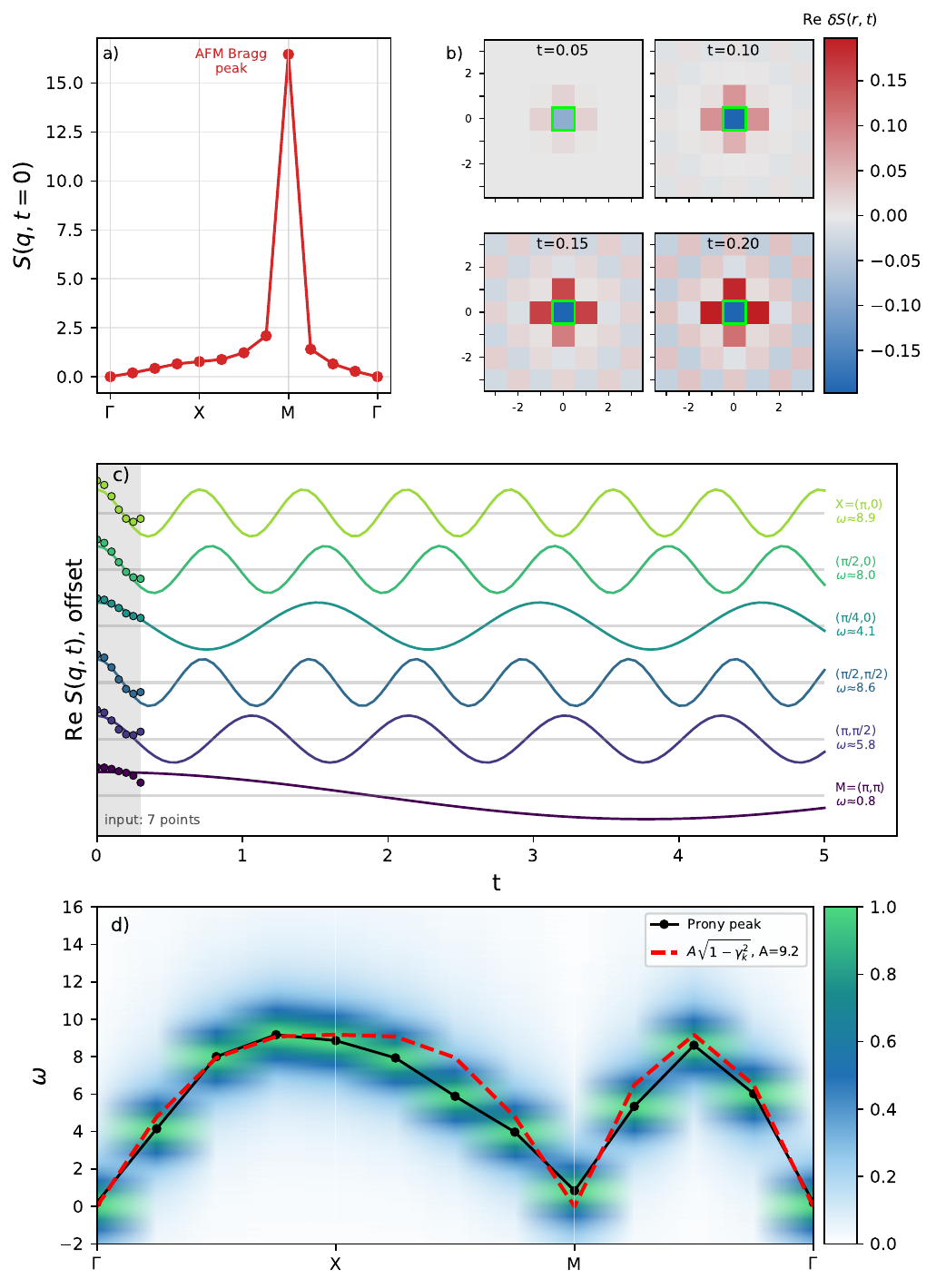}
    \caption{2D Heisenberg excitation spectrum from the improved MPS ground
    state ($8\times8$, coefficient cutoff $10^{-8}$, weight cutoff $w=4$, clip
    $\varepsilon_{\mathrm{clip}}=10^{-3}$, bond dimension $\Dchi=800$), mirroring Fig.~\ref{fig:2D_AF}.
    a) Static structure factor $S(k,t=0)$; the larger antiferromagnetic Bragg
    peak at $M$ reflects the more accurate order parameter of the higher-$\Dchi$
    ground state. b) Spreading of the excitation $\delta S(r,t)=S(r,t)-S(r,0)$
    for the time points used. c) Early-time data and the rank-1 PSD extension for
    selected momenta (normalized per $q$ point). d) Spin structure factor
    $|S(k,\omega)|^2$ from the extended spectra, extracted frequencies from the
    rank-1 decomposition, and a fit to the linear spin-wave spectrum
    $\omega_q = A\sqrt{1-\gamma_k^2}$.}
    \label{fig:2D_AF_improved_arnab}
\end{figure}

\begin{figure}[b]
    \includegraphics[width=0.49\textwidth]{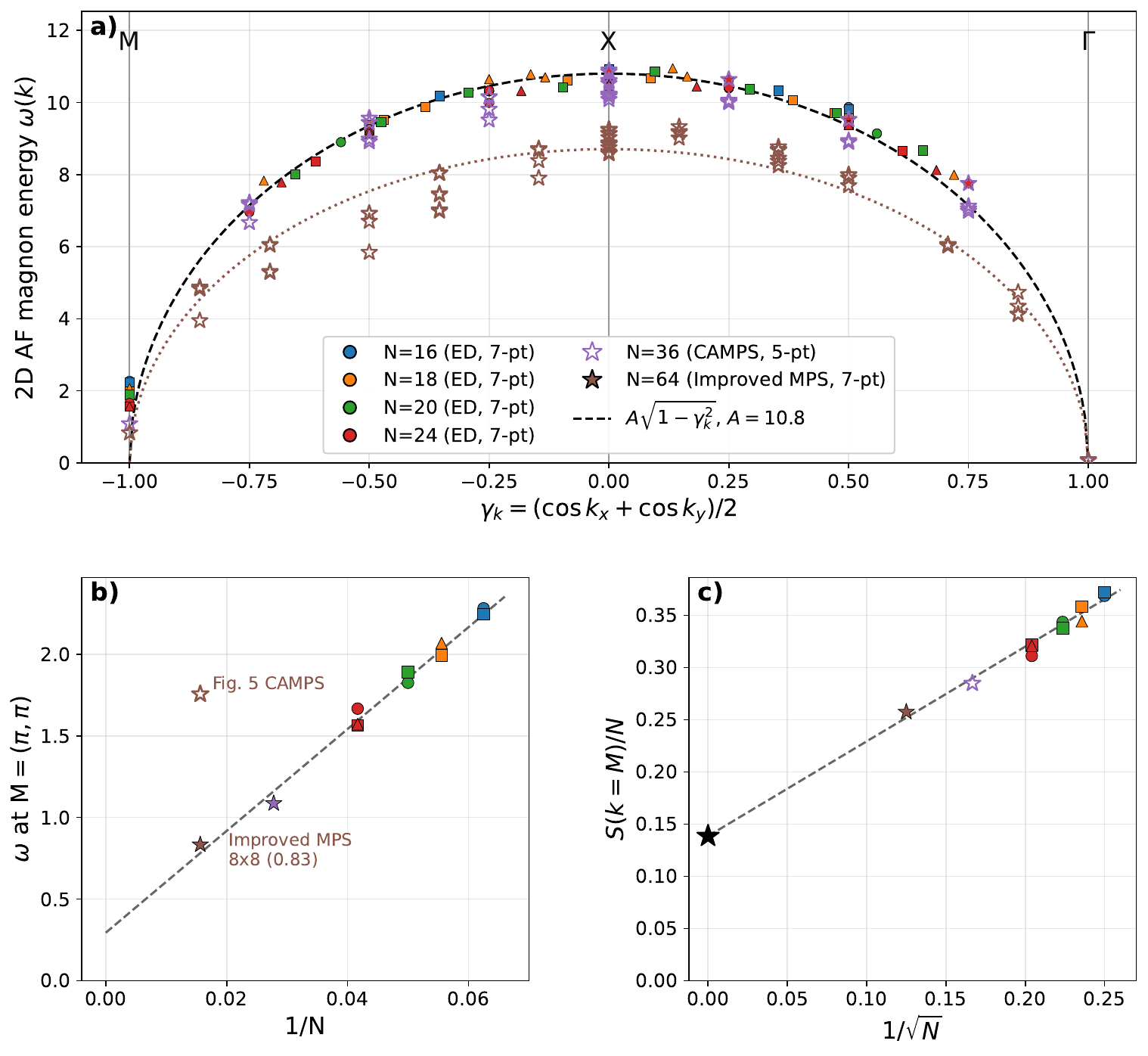}
    \caption{Finite-size analysis of the 2D anti-ferromagnetic Heisenberg model.
    Colors denote system size $N$; the different marker shapes of a given color are
    inequivalent Betts clusters~\cite{Betts:1999} of that size (exact diagonalization,
    $N=16$--$24$), and stars are tensor-network grid clusters ($6\times6$ CAMPS, open;
    improved $8\times8$ MPS, filled). Frequencies are extracted from a rank-1 Prony fit
    to a $7$-point short-time window.
    a)~Cross-cluster collapse of the magnon dispersion $\omega(k)$ onto the linear-spin-wave arch
    $A\sqrt{1-\gamma_k^2}$ (dashed: fit to the ED clusters; dotted: the
    finite-size-narrowed improved $8\times8$).
    b)~Finite-size scaling of the soft-mode energy at $M=(\pi,\pi)$; the filled star is the improved $8\times8$ MPS and the
    open star the main-text $8\times8$ data of Fig.~\ref{fig:2D_AF}.
    c)~Static staggered structure factor $S(k=M)/N$; the finite
    intercept (star) gives an ordered moment $m_s\approx0.32$, close to the exact
    value $0.31$, signaling long-range N\'eel order.}
    \label{fig:afm_fss}
\end{figure}

Figure~\ref{fig:2D_AF_improved_arnab} illustrates this improved scheme, using
the $(w,\varepsilon)=(4,10^{-8})$ evolution-truncation combination for the single-site Pauli evolution, clipped at $\varepsilon_{\rm clip}=1\times10^{-3}$ before conversion to an MPO, and
contracted against the $D=800$ truncated MPS to obtain the correlation
function in the time domain.
Panel (a) shows the static structure factor $S(k,t=0)$: the antiferromagnetic Bragg peak at $M=(\pi,\pi)$ is larger and better resolved than in Fig.~\ref{fig:2D_AF}, reflecting the more accurate order parameter obtained slightly improved ground state
and more strict PP parameters i.e., $(w,\varepsilon)=(4,10^{-8})$ compared to 
$(w,\varepsilon)=(5,10^{-4})$ used in Fig.~\ref{fig:2D_AF}. 
Panels (b)--(c) show the resulting real-space spreading of the perturbation and
the early-time correlation data together with its rank-1 PSD extension.
Panel (d) shows the extracted dynamical structure factor $|S(k,\omega)|^2$ together with a fit to the linear spin-wave dispersion $\omega_q=A\sqrt{1-\gamma_k^2}$; relative to Fig.~\ref{fig:2D_AF}, the spurious softening of the magnon mode near $M$ is reduced, giving closer agreement with the expected spin-wave spectrum across the Brillouin zone.

The improved MPS state and Pauli propagation truncations used aid in moving the results closer to the expected, including the finite size effects. We demonstrate this in Fig.~\ref{fig:afm_fss} by comparing to smaller finite size exact diagonalization studies on bipartite Betts clusters~\cite{Betts:1999}---compact tilings of the square lattice with $N=16,18,20,24$ sites and several inequivalent geometries at each size---for which the ground state and the short-time correlator (7 points) can be computed exactly. For every allowed momentum $k$ we extract the magnon frequency $\omega(k)$ from a rank-1 Prony fit to a matched short-time window; we also evaluate the static structure factor $S(k=M)$ at the ordering wavevector $M=(\pi,\pi)$.

Figure~\ref{fig:afm_fss}(a) shows that $\omega(k)$ from every cluster, independent of size and geometry, collapses onto a single linear-spin-wave arch $\omega(k)=A\sqrt{1-\gamma_k^2}$, confirming that these small tilings reproduce the thermodynamic dispersion away from the ordering vector. The improved $8\times8$ data (stars) lie systematically below the small-cluster arch; because the Betts data are exact, this likely comes from a combination of a decreased finite-size effect and errors introduced by the Pauli propagation. Panels (b) and (c) examine the ordering wavevector directly. The dynamical soft-mode energy $\omega_M$ decreases monotonically with system size [panel (b)], consistent with the Anderson tower of states~\cite{Anderson:1952} that collapses onto the symmetry-broken ground state as $\sim 1/N$. At the same time the static order parameter $S(k=M)/N$ stays finite [panel (c)], extrapolating with the expected $1/\sqrt N$ correction~\cite{Huse:1988} to $S(k=M)/N\to0.14$, i.e.\ a staggered moment $m_s\approx0.32$, in good agreement with the quantum Monte Carlo value $0.307$~\cite{Sandvik:1997}. The finite extrapolated static order parameter indicates that the Goldstone mode at $S(k=M)$ should go to zero in the thermodynamic limit; the small residual $\omega_M$ at the largest sizes is limited by the short extraction window rather than by a physical gap.

\begin{figure}[t]
    \vspace{0.1in}
    \centering
    \includegraphics[width=0.95\linewidth,clip=true, trim=0 0 0 0]{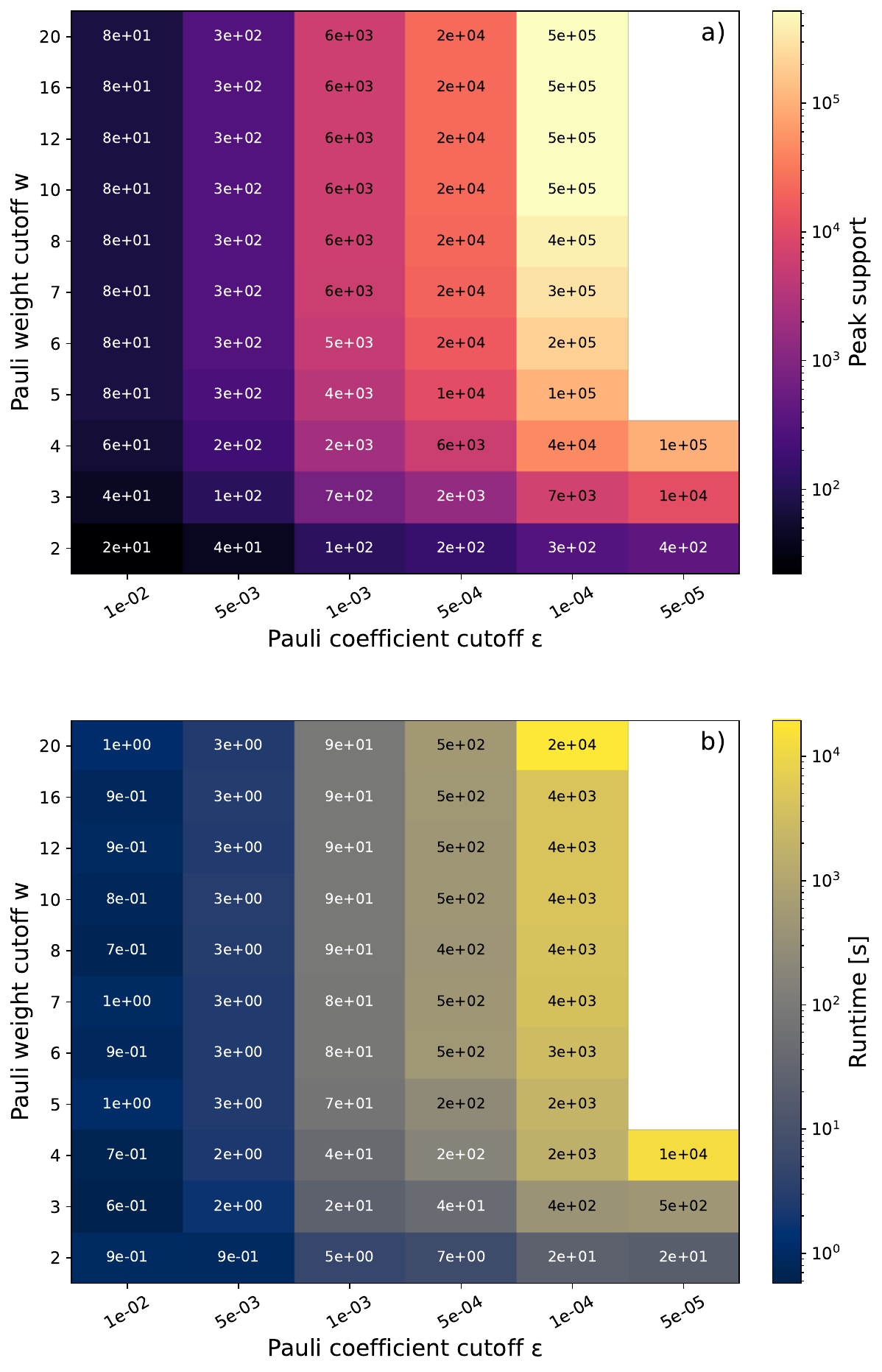}
    \caption{Truncation cost on the same $N=64$ FM Heisenberg
    setup as Fig.~\ref{fig:n64_fm_overview}, as a function of Pauli
    coefficient cutoff $\varepsilon$ and Pauli weight cutoff
    $w$. a) Peak Pauli-string count during the
    trajectory ($t \in [0, 10]$). b) Cumulative
    Trotter-step time per cell. Note that the full space of Pauli strings for
    this system size is $\mathcal{O}(10^{38})$.}
    \label{fig:n64_fm_cost_heatmap}
\end{figure}

\section{Computational cost of Pauli propagation for the Heisenberg model}
\label{app:walltime}

Fig.~\ref{fig:n64_fm_cost_heatmap} shows the maximal Pauli weight as well as the wall time associated with performing the Pauli propagation calculation for the 64-site Heisenberg model.  The calculations were performed using an M5 Pro cpu.  As expected, the peak operator support as well as the wall time increase markedly as the Pauli propagation calculation gets better (smaller $\varepsilon$, larger weight cutoff $w$).  Note that the dimension of the full operator space for 64 is
$4^{64}-1\approx 3.4 \times 10^{38}$; the relatively severe truncation limits the number of Pauli terms by $30+$ orders of magnitude.

%\clearpage

\section{Ferromagnetic spin waves on large lattices}
\label{app:largeFM}

\begin{figure}[htpb]
    \centering
    \includegraphics[width=0.49\textwidth]{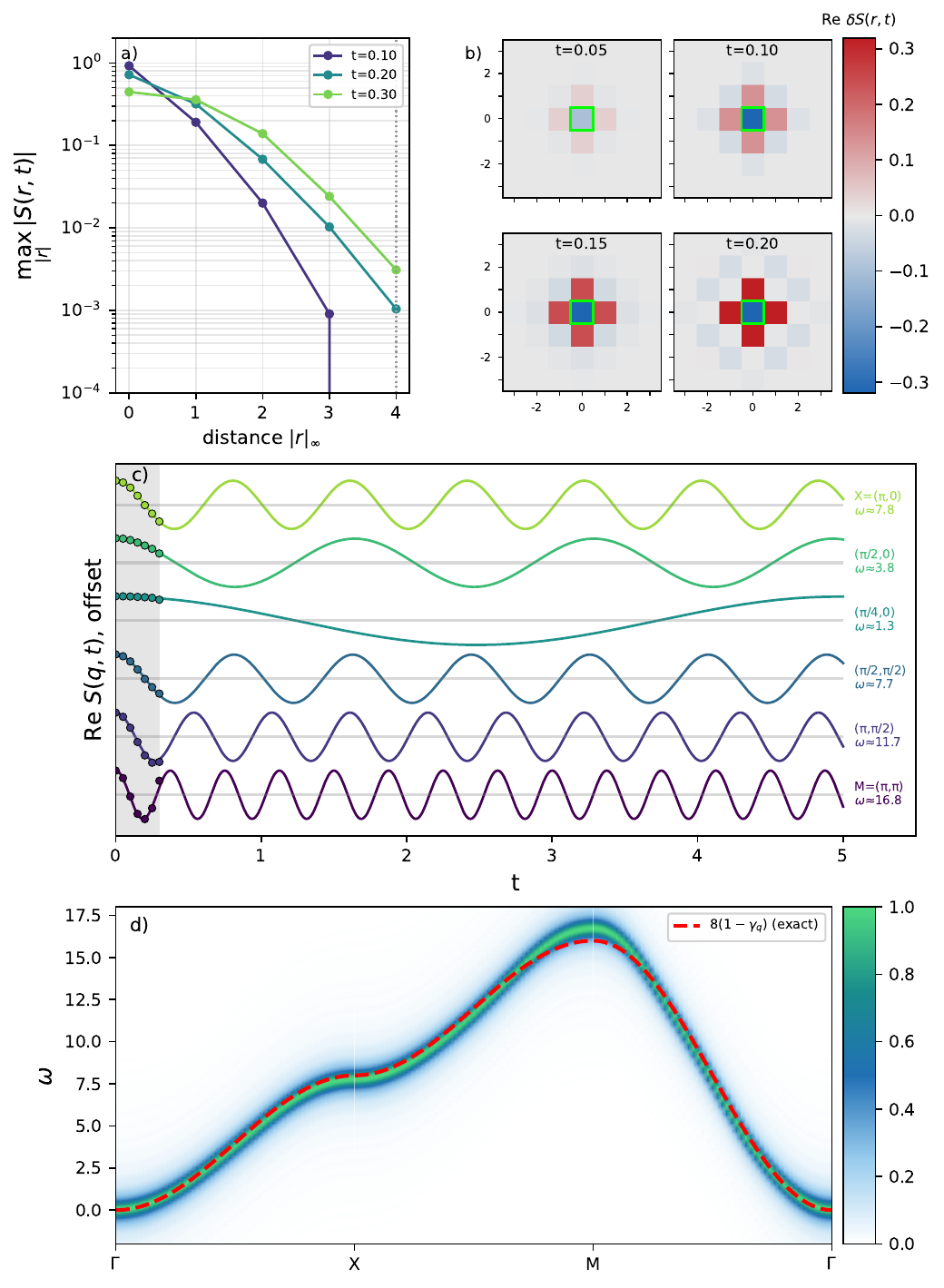}
    \caption{Ferromagnetic spin waves on a $100\times100$ lattice reconstructed from an $8\times8$ Pauli patch. (a) Locality: $\max_{|r|}|S(r,t)|$ as a function of distance $|r|_\infty$ (dotted line: patch half-width), showing the correlator is confined within the patch. (b) Real-space spread $\operatorname{Re}\delta S(r,t)$ of the excitation. (c) Short-time data and the rank-1 PSD extension for selected momenta. (d) $S(k,\omega)$ on a $100\times100$ momentum grid obtained by embedding the patch correlator, compared with the exact linear spin wave $\omega_q = 8(1-\gamma_k)$ (red dashed).}
    \label{fig:largeFM}
\end{figure}

For a ferromagnet the ground state is a fully polarized product state, so evaluating a Pauli-propagated operator in it is a trivial contraction---only single-spin-flip strings survive---with no ground-state preparation cost. Together with the short-time locality of the operator, this means the correlation function on a large lattice can be reconstructed from a small patch. Figure~\ref{fig:largeFM}(a) shows that $S(r,t)$ stays confined within an $8\times8$ patch: even at the latest time used, the signal has dropped by more than two orders of magnitude at the patch boundary. Because $S(r,t)$ is translation invariant and locally supported, the $8\times8$ patch reproduces the thermodynamic-limit correlator, which we embed into a $100\times100$ real-space grid and Fourier transform to arbitrary momentum resolution [panels b)--d)]. The rank-1 PSD extension recovers the exact spin-wave dispersion $\omega_q = 8(1-\gamma_k)$ across the entire Brillouin zone, gapless at $\Gamma$ and peaking at $M$. 

\end{document}